\documentclass[twocolumn]{aastex63}

\received{-}
\revised{-}
\accepted{-}
\submitjournal{ApJL}

\shorttitle{Altair's Radio Spectrum}
\shortauthors{White et al.}
\graphicspath{{./}{figures/}}

\begin{document}

\title{The first radio spectrum of a rapidly rotating A-type star}

\correspondingauthor{Jacob Aaron White}
\email{jwhite@nrao.edu}

\author[0000-0001-8445-0444]{Jacob Aaron White}
\affiliation{National Radio Astronomy Observatory, 520 Edgemont Rd., Charlottesville, VA, 22903, USA}
\affiliation{Jansky Fellow of the National Radio Astronomy Observatory}

\author[0000-0001-9132-7196]{F. Tapia-V\'azquez}
\affiliation{Instituto de Radioastronom\'\i a y Astrof\'\i sica, Universidad Nacional Aut\'onoma de M\'exico, P.O. Box 3-72, 58090, Morelia, Michoac\'an, M\'exico.}
\affiliation{GICC, Interdisciplinary Group of Scientific Computing, UNAM}

\author[0000-0002-3446-0289]{A.~G.~Hughes}
\affil{Department of Physics and Astronomy,
University of British Columbia,
6224 Agricultural Rd.,
Vancouver, BC V6T 1T7, Canada}

\author{A. Mo\'or}
  \affiliation{Konkoly Observatory,
  Research Centre for Astronomy and Earth Sciences,
  E\"otv\"os Lor\'and Research Network (ELKH),
  Konkoly-Thege Mikl\'os \'ut 15-17, 1121 Budapest, Hungary}
\affiliation{ELTE E\"otv\"os Lor\'and University, Institute of Physics, P\'azm\'any P\'eter s\'et\'any 1/A, 1117 Budapest, Hungary}

\author[0000-0003-3017-9577]{B. Matthews}
\affiliation{Herzberg Institute,
 National Research Council of Canada,
 5071 W. Saanich Road,
 Victoria, BC V9E 2E7, Canada}

\author[0000-0003-1526-7587]{D. Wilner}
\affiliation{Center for Astrophysics | Harvard \& Smithsonian,
60 Garden Street,
Cambridge, MA 02138, USA}

\author{J. Aufdenberg}
\affiliation{Physical Sciences Department,
Embry-Riddle Aeronautical University,
600 S Clyde Morris Blvd.,
Daytona Beach, FL 32114, USA}

\author[0000-0003-3453-4775]{O. Feh\'er}
  \affiliation{IRAM, 300 Rue de la piscine, 38406 Saint-Martin-d’H\`eres, France}

\author{A. M. Hughes}
\affiliation{Department of Astronomy,
Van Vleck Observatory,
Wesleyan University,
96 Foss Hill Dr.,
Middletown, CT 06459, USA}

\author[0000-0003-0257-4158]{V. De la Luz}
\affiliation{Escuela Nacional de Estudios Superiores Unidad Morelia, 
Universidad Nacional Aut\`onoma de M\'exico, 
Morelia, 58190, M\'exico}
\affiliation{GICC, Interdisciplinary Group of Scientific Computing, UNAM}

\author[0000-0002-4146-7921]{A. McNaughton}
\affiliation{Physical Sciences Department,
Embry-Riddle Aeronautical University,
600 S Clyde Morris Blvd.,
Daytona Beach, FL 32114, USA}
\affiliation{Earth \& Biological Sciences Directorate,
Biological Sciences Division,
Pacific Northwest National Laboratory, Richland, Washington 99352, USA}

\author[0000-0003-2343-7937]{L. A. Zapata}
\affiliation{Instituto de Radioastronom\'\i a y Astrof\'\i sica, Universidad Nacional Aut\'onoma de M\'exico, P.O. Box 3-72, 58090, Morelia, Michoac\'an, M\'exico.}



\begin{abstract}

The radio spectra of main-sequence stars remain largely unconstrained due to the lack of observational data to inform stellar atmosphere models. As such, the dominant emission mechanisms at long wavelengths, how they vary with spectral type, and how much they contribute to the expected brightness at a given radio wavelength are still relatively unknown for most spectral types. We present radio continuum observations of Altair, a rapidly rotating A-type star. We observed Altair with NOEMA in 2018 and 2019 at 1.34 mm, 2.09 mm, and 3.22 mm and with the VLA in 2019 at 6.7 mm and 9.1 mm. In the radio spectra, we see a brightness temperature minimum at millimeter wavelengths followed by a steep rise to temperatures larger than the optical photosphere, behavior that is unexpected for A-type stars. We use these data to produce the first sub-millimeter to centimeter spectrum of a rapidly rotating A-type star informed by observations. We generated both PHOENIX and KINICH-PAKAL model atmospheres and determine the KINICH-PAKAL model better reproduces Altair's radio spectrum. The synthetic spectrum shows a millimeter brightness temperature minimum followed by significant emission over that of the photosphere at centimeter wavelengths. Together, these data and models show how the radio spectrum of an A-type star can reveal the presence of a chromosphere, likely induced by rapid rotation, and that a Rayleigh Jean's extrapolation of the stellar photosphere is not an adequate representation of a star's radio spectrum. 

\end{abstract}

\keywords{stars: individual (Altair) - radio continuum: stars - stars: atmospheres - submillimeter: stars - millimeter: stars - circumstellar matter}


\section{Introduction} \label{sec:intro}

The internal structure of main-sequence stars, and thus the mechanism(s) for generating magnetic fields, is largely dependent on mass and therefore spectral type \citep{charbonneau13}. 
M-type stars can be fully convective and generate strong magnetic fields through an $\alpha^{2}$ dynamo \citep[e.g.,][]{chabrier06}. K through early F-type stars have radiative cores with convective outer layers, and the shear between these zones is thought to play a critical role in magnetic field generation \citep[e.g.,][]{guerrero16}. A-type and more massive stars are mostly radiative and unable to generate strong magnetic fields. \citet{shorlin02} confirm through a spectropolarimetric survey that main-sequence ``normal" A-type stars do not have significant magnetic fields. Some A-type stars, however, and in particular those that are chemically peculiar, have been observed to have non-negligible magnetic field strengths \citep[see, e.g.,][]{preston74}.

Stellar magnetic fields and convective zones can heat the upper atmosphere and are therefore associated with chromospheric and coronal activity \citep{ayres}. For these stars, the radio emission is proportional to the plasma temperature ($\rm T_{R}$) in the atmosphere and is dominated by optically thick free-free radiation from electrons, ions, and H$^-$ \citep{liseau13, linsky17}. Therefore, a brightness temperature ($\rm T_{B}$) spectrum can trace the temperature and density structure. Radio observations with broad spectral coverage provide a reliable method of measuring the thermal emission as a function of height above the stellar surface and modeling the thermal structure of the atmosphere \citep[e.g., the KINICH-PAKAL (KP) modeling code][]{tapia20}. Higher atmospheric temperatures in the chromosphere can lead to a higher $\rm T_{B}$ at radio wavelengths, and are evidenced by a rise in the spectrum at wavelengths $>1$ mm \citep{loukitcheva04, white20}.

For solar-like stars, the sub-millimeter (submm) emission is expected to arise from the photosphere \citep{liseau13}. At $\sim$mm wavelengths, the emission is generated in both the upper photosphere and lower chromosphere. At mm-cm wavelengths, the emission is expected to arise from the lower to the upper chromosphere \citep{Liseau}. At $>$cm wavelengths, the emission comes from the upper chromosphere and lower corona \citep{trigilio}.

A-type stars lack significant stellar convectivity, meaning chromosphere and corona are not expected. This in principle leads to a thermal submm-cm spectrum largely representative of the optical photosphere in the Rayleigh-Jeans limit. For Sirius A, a slowly rotating A0V star, radio observations \citep{white18,white19} and PHOENIX photosphere models \citep{hauschildt99} show a spectrum slightly cooler than the photosphere and no evidence of an increase in $\rm T_{B}$ that could be attributed to a chromosphere or corona (for details on PHOENIX and KINICH-PAKAL modeling codes, see Sec.\,\ref{sec:model}).

In general, stellar spectra in the submm-cm wavelength regime are difficult to accurately model due to the lack of empirical data in this portion of the spectrum which can be used to inform existing models. The dominant atmospheric processes and structure that determine the spectral shape rely strongly on intrinsic stellar properties. While a model of a single spectral type may be applicable to a few stars, it may significantly over- or under-predict the flux of many other stars, depending on the wavelength. 

In this letter, we present the first sub-millimeter to centimeter spectrum and model of Altair ($\alpha$ Aql, HD 187642). This 1.3 Gyr A7IV-V star is one of the most studied objects in our solar neighborhood \citep[$5.13\pm0.01$ pc;][]{leeuwen07}. Altair is known for low amplitude pulsation \citep[$\delta$ Scuti classification,][]{buzasi05}, rapid rotation \citep[$\rm v\,sin(i) =190-250~km~s^{-1}$,][]{abt95}, and significant oblateness \citep[e.g.,][CHARA measures a polar radius of $1.63~\rm R_{\odot}$ and an equatorial radius of $2.03~\rm R_{\odot}$]{monnier07}. This oblateness has contributed to a $6860 - 8450$ K temperature gradient from the equator to the poles \citep[surface averaged temperature of 7594 K;][]{bouchaud20}, although it is unclear if this temperature gradient is present at longer wavelengths. Altair's rapid rotation has created a stable equatorial dynamo, which generates stronger than expected magnetic fields for an A-type star \citep{robrade09}. In addition, UV spectroscopy shows evidence of significant heating in Altair's upper atmosphere \citep{walter95} which is interpreted by some as chromospheric emission \citep{simon95}. This hotter upper atmosphere may cause significant deviations from an assumed thermal spectrum at long wavelengths as well.

These results are part of The MESAS Project (Measuring the Emission of Stellar Atmospheres at Sub-millimeter to centimeter wavelengths), an ongoing observational campaign that seeks to build a catalog of radio spectra of main-sequence stars. This project uses long wavelength data to inform stellar atmosphere modeling codes in order to better understand the dominant stellar emission mechanisms at these wavelengths.

\section{Observations} \label{sec:obs}

\begin{figure*}
\centering
\includegraphics[width=\textwidth]{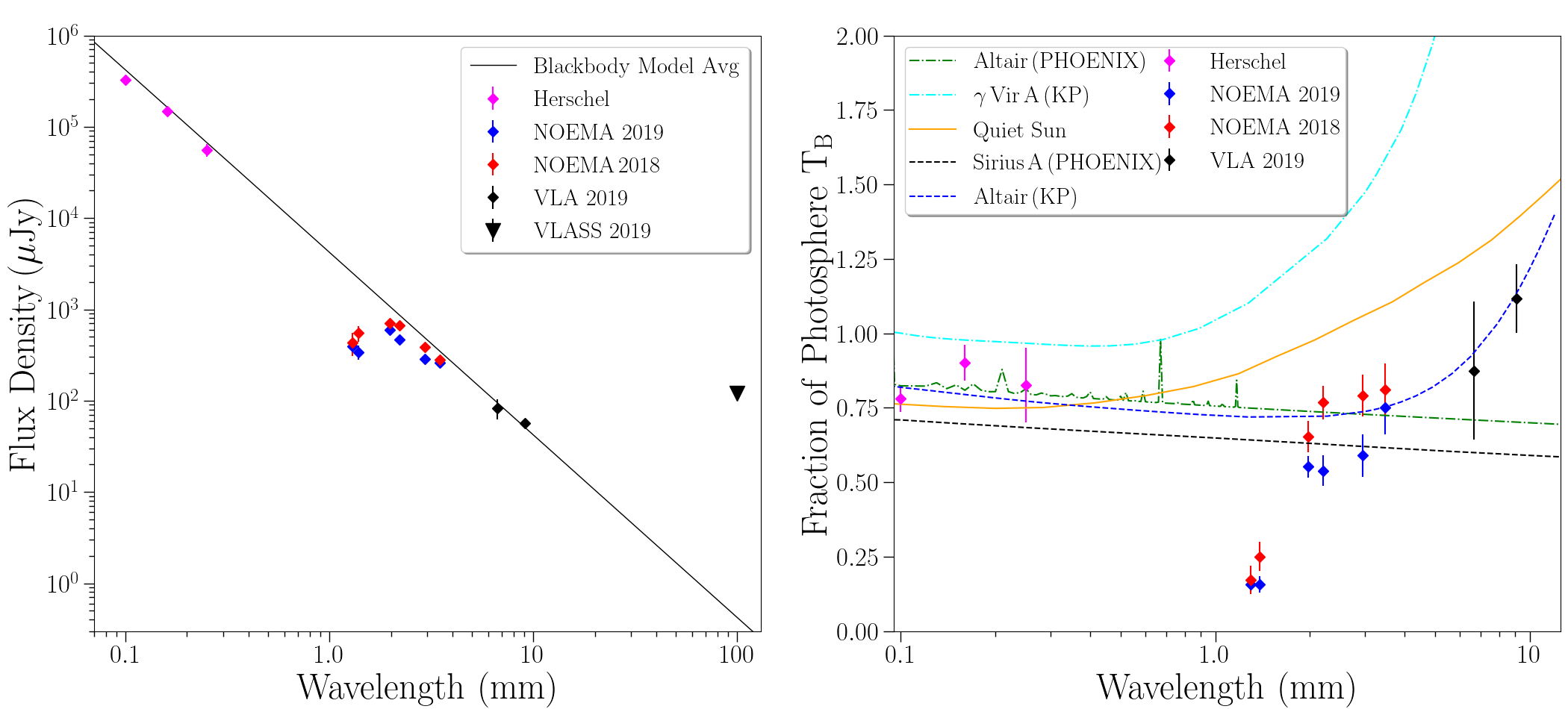}
\caption{\textbf{Left:} Altair's spectrum. The data from Table \ref{data_summary} are plotted along with the VLASS $\sigma_{rms}$ from the non-detection at 100 mm (3 GHz) included for illustrative purposes. The solid black line is a 7594 K blackbody model and is included to show the significant deviations at different wavelengths. 
\textbf{Right:} Altair's fractional $\rm T_{B}$ spectrum and different stellar emission models. The y-axis is $\rm T_{B}$ divided by the optical photosphere temperature (where 1.0 represents pure blackbody emission). The cyan line is a KP model \citep{tapia20} of the F0IV star $\gamma$ Vir A \citep{white20}. The orange line is a quiet Sun model \citep{loukitcheva04}.  The black line is a PHOENIX model \citep{hauschildt99} of the A0V star Sirius A \citep{white18,white19}. The blue and green lines are KP and PHOENIX models of Altair (see Sec.\,\ref{sec:model}). The PHOENIX and KP models are detailed in Sec.\ref{sec:model}.} \label{tb_plot}

\end{figure*}

The analysis in this letter uses new observations and archival data from \textit{Herschel} and Very Large Array Sky Survey (VLASS). The observations and calibration procedures for the NOrthern Extended Millimeter Array (NOEMA) and NSF's Karl G. Jansky Very Large Array (VLA) data are described in Appendix \,\ref{data_app}. The primary observational results are summarized in Table \ref{data_summary}. 

Fig.\ref{tb_plot} shows Altair's radio spectrum. This spectrum clearly deviates from a blackbody (black line) at multiple wavelengths and such a model will over- or under-predict the flux depending on the wavelength. On the right side of Fig.\ref{tb_plot}, we show Altair's spectrum as a fraction of the photosphere temperature. We calculate the brightness temperature assuming an elliptical emitting area using the polar and equatorial radii. In this spectral representation, 1.0 indicates an observed radio $\rm T_{B}$ equal to the surface averaged optical photosphere temperature of 7594 K. This figure allows for a straightforward comparison of different spectral types.  In addition to Altair models, Fig.\,\ref{tb_plot} includes a KP model of $\gamma$ Vir A \citep{white20}, a quiet Sun model \citep{loukitcheva04}, and a PHOENIX model of Sirius A \citep{white18, white19}. The PHOENIX and KP models are detailed in Sec.\,\ref{sec:model}.

\section{Discussion} \label{sec:disc}

Data-informed atmosphere models of solar-type stars, such as the Sun and $\gamma$ Vir A, have a shallow $\rm T_{B}$ minimum below 1 mm and a large increase in $\rm T_{B}$ at wavelengths over 1 mm. This increase is due to the presence of a stellar chromosphere and corona, which contribute to significant emission over the photosphere \citep{loukitcheva04, white20}. Altair appears to exhibit a very strong $\rm T_{B}$ minimum, dropping to at least $0.2\times T_{phot}$.  The representative wavelength of the minimum is unknown due to a lack of  $0.25-1.29$ mm data. Regardless, from $1.29-3.48$ mm the $\rm T_{B}$ rapidly rises back to near the photosphere temperature ($\sim 0.80\times T_{phot}$). At longer wavelengths, the rise continues.

As Altair is an A7V star, and data-informed atmosphere models of A-type stars are largely non-existent in the literature, it may have been naively assumed that Altair's radio spectrum would be similar to that of Sirius A (an A0V star; included in Fig.\,\ref{tb_plot}). Altair's spectrum, however, significantly deviates from a Sirius-like profile. Since Altair is rapidly rotating, leading to an equatorial magnetic dynamo, the possibility exists that its spectrum would be more similar to magnetically active stars such as the Sun with significant emission in the chromosphere \citep[e.g.,][]{ayres}. Comparing Altair's spectrum to a quiet Sun model \citep[e.g., the orange line on right-hand-side of Fig.\,\ref{tb_plot};][]{loukitcheva04} we see that a Solar-like model may be more representative than a Sirius-like model, but still does not adequately explain the spectrum.

\subsection{Spectral Index}

\begin{figure}
\centering
\includegraphics[width=0.5\textwidth]{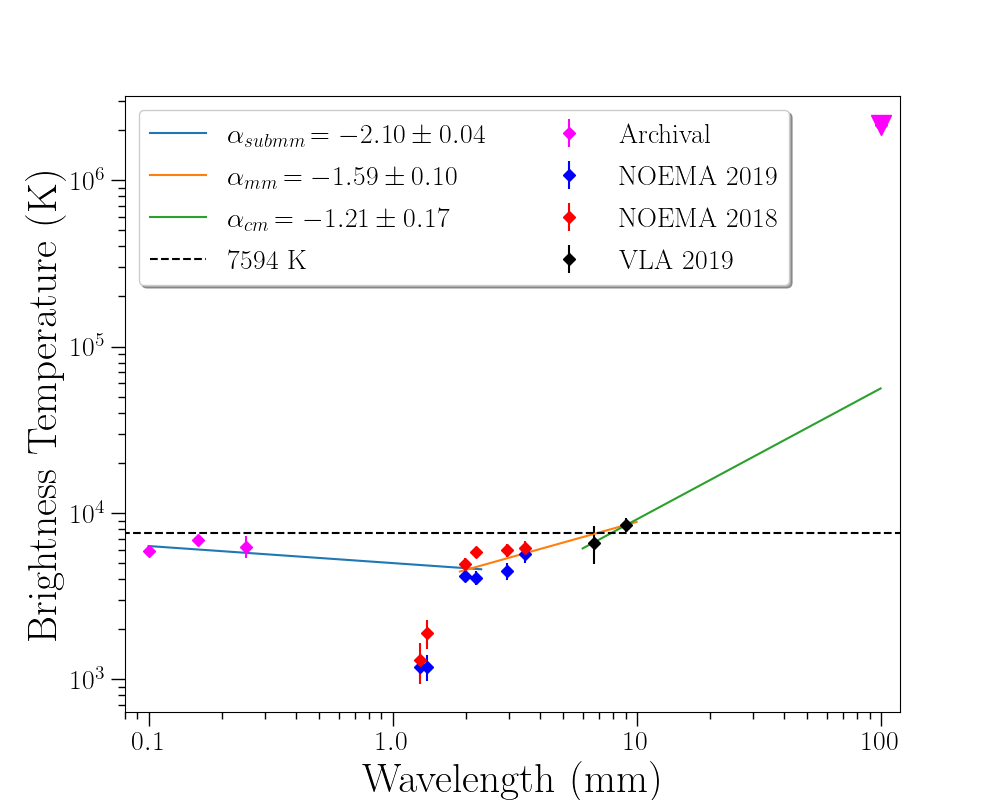}
\caption{Altair's $\rm T_{B}$ spectrum. The \textit{Herschel} data are plotted in purple, the VLA data in black, and the 2018 and 2019 NOEMA data in red and blue, respectively. The $\sigma_{rms}$ of the VLASS non-detection is denoted by the upside down triangle. The 7594 K average optical photosphere temperature (black line) represents a spectral index of $\alpha=-2.0$. The submm spectral index (blue line) is $\alpha_{submm}=-2.10\pm0.04$. This fit is to only the \textit{Herschel} and NOEMA $2\,\rm mm$ data. The millimeter spectral index (orange line) is $\alpha_{mm}=-1.59\pm0.10$. This fits from the NOEMA 2 mm data to the VLA 9 mm data. The cm spectral index (green line) is $\alpha_{cm}=-1.21\pm0.17$ and only fits the VLA data.}\label{spectral_index}
\end{figure}

A given stellar emission mechanism will have a characteristic spectral index, $\alpha$, related to the observed flux by $F_{\lambda}\propto \lambda^{\alpha}$.  Therefore, a spectral index can be a powerful diagnostic tool for disentangling emission mechanisms and can provide information about the structure of a star's atmosphere. Thermal emission from an optically thick source will have a spectral index of $\alpha = -2$. Given a lack of observational data at long wavelengths, this values of $\alpha$ is commonly assumed to be the representative for A-type stars in the Rayleigh-Jeans limit and is used to, e.g., estimate the stellar flux contribution in circumstellar disk studies (see Sec.\,\ref{sec:disks}). Since Altair's spectrum is clearly not well represented by a blackbody model, then there are likely additional emission/absorption mechanisms contributing to its spectrum.

In Fig.\,\ref{spectral_index}, we show Altair's spectrum along with spectral indices fit to subsets of the data using a least squares approach with the {\scriptsize SciPy} function \textit{curve\_fit} \citep{virtanen20}. For $\lambda <$ 1.98 mm, we calculate a spectral index of $\alpha_{submm}=-2.10 \pm 0.04$. Within the $1\sigma$ uncertainties, this is consistent with a deviation from the thermal photospheric emission in the Rayleigh-Jeans regime. A similar cooler spectrum with a slightly steeper spectral index was observed in Sirius A and is well-reproduced by a PHOENIX atmosphere model \citep[black line in Fig.\,1][]{white18, white19}. We do not include the 1.3 mm data in the spectral index calculations as the drop in $\rm T_{B}$ to $1000-2000$ K is not reconcilable with any known physical mechanism (see Sec.\,\ref{sec:model}).

For wavelengths in the range of 1.98 mm $< \lambda <$  10 mm, the spectral index has a value of $\alpha_{mm} =-1.59 \pm 0.10$. In this portion of the spectrum, the spectral index is in broad agreement with solar-like stars \citep[e.g.,][]{villadsen,Liseau}, where chromospheric emission begins to contribute significantly to the observed flux. Considering only the VLA data, we find $\alpha_{cm} = -1.21 \pm 0.17$, continuing the trend towards higher brightness temperatures and a possible contribution from chromospheric emission. Due to the lack of data at wavelengths longer than 1 cm we are unable to determine if the cm spectral index is also solar-like.

\subsection{Stellar Atmosphere Models} \label{sec:model}

\begin{figure*}
\centering
\includegraphics[width=\textwidth]{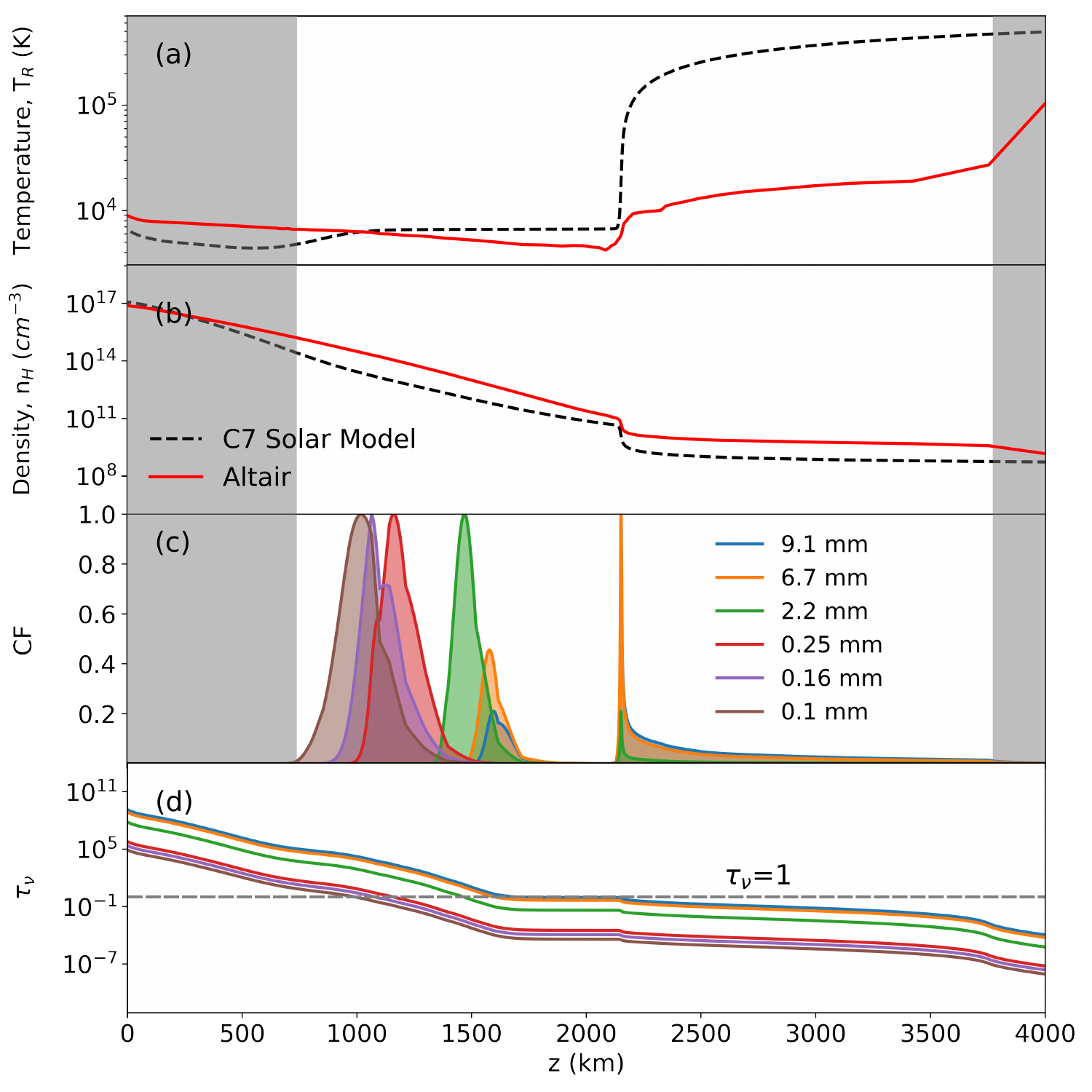}
\caption{KP model of Altair's atmosphere as a function of height above the photosphere ($z=0$ at $\tau_{\rm 500\,nm}=1$). The shaded regions represent the outer boundaries where KP cannot be accurately applied. (a) The red line shows Altair's $\rm T_{R}$ profile. The black line is the C7 Solar model \citep{avrett08} used as initial conditions. (b) Density profile of Altair (red line) and the average solar values (black line). (c) Normalized contribution function (CF) for Altair. At 9.1 mm, the maximum contribution occurs at 2152 km and a second peak at 1596 km has a contribution of 21\%. At 6.7 mm, the CF has a maximum at 2152 km and a second peak at 1578 km with a contribution of 45\%. At 2.2 mm, the maximum is reached at 1470 km and a second peak at 2152 km with a contribution of 21\%. At 0.25 mm, 0.16 mm, and 0.10 mm, the CF shows peaks at 1162 km, 1065 km, and 1019 km, respectively. (d) The optical depth at each wavelength. \label{kp-altair} }
\end{figure*}

To model Altair's spectrum, we consider both PHOENIX \citep{hauschildt99} and KP atmospheres \citep{tapia20}. The PHOENIX model includes only the stellar photosphere whereas the KP model includes chromospheric emission from above the photosphere. The MESAS Project previously used PHOENIX to model Sirius A \citep{white18, white19} and KP to model $\gamma$ Lep, $\gamma$ Vir A, and $\gamma$ Vir B \citep{white20}.

The PHOENIX approach (version 17.01.02A) includes a gravity-darkened model for Altair's photosphere. We adopted Altair's `$\beta$-fixed' parameters \citep{monnier07} and a $0\farcs19495$ parallax \citep{leeuwen07}. We used the modeling code from \citet{aufdenberg06} which was previously used to model the photosphere of the rapid-rotator Vega \citep[a similar approach is also used in][]{lipatov20}. We used 1-D PHOENIX model atmospheres to generate 117 LTE radiation fields evaluated at 78 different angles, $T_{\rm eff}$ values from $6750 - 8750$ K in 250 K intervals, and $\log_{10}(g)$ values from $3.90-4.40$ in 0.05 dex intervals. The temperature-pressure structure of each model atmosphere in radiative equilibrium was computed assuming a hydrostatic atmosphere, including between 1719584 and 1842901 spectral lines in LTE, depending on the T$_{eff}$ and $\log_{10}(g)$ values, and 576 bound-bound and bound-free transitions from hydrogen and helium in non-LTE. We then sampled the oblate surface with 100 latitude points and 300 longitude points, integrating the intensities to produce a synthetic photosphere spectrum (see Fig.\,\ref{tb_plot}).

KP is a non-linear framework that computes a stellar atmosphere spectrum through a 1-D NLTE semi-empirical model of the chromosphere at long wavelengths. KP uses PakalMPI \citep{delaluz2011} to generate an atmosphere structure in hydrostatic equilibrium and to compute a synthetic spectrum taking into account three opacity functions (Bremsstrahlung, H$^{-}$, and inverse Bremsstrahlung). A Levenberg–Marquardt algorithm is used to iteratively modify the $\rm T_{R}$ profile, fitting the synthetic spectrum to the observed radio fluxes. KP computed 11848 atmospheres to obtain a new semi-empirical model of Altair. The last radial $\rm T_{R}$ profile is then smoothed using a Savitzky–Golay filter with a seventh order polynomial. This model is equilibrated again with PakalMPI to guarantee hydrostatic equilibrium. The final model structure is shown in Figs.\,\ref{kp-altair}(a)–(b).

We determine that the KP model better reproduces the data and include the final semi-empirical model in Fig.\,\ref{kp-altair}. This model shows that for wavelengths $<1$ mm the emission comes from the photosphere and for $>3$ mm the emission originates in the chromosphere. This is due to an optically thick atmosphere at altitudes where $\rm T_{R}$ is below 10000 K and the atmosphere is not fully ionized. At 2083 km above the photosphere, we find a $\rm T_{R}$ minimum of 4207 K (T$_{R,min}$/T$_{eff}\approx0.55$) and a hydrogen density of $\rm n_H=1.5 \times 10^{11}~cm^{-3}$. Overall, $\rm T_{R}$ gradually decreases until reaching a minimum, typical of photospheric behavior. Starting at a height of 2083 km, we observe a positive temperature gradient that extends for $\sim100$ km. $\rm T_{R}$ then plateaus at 10000 K and extends $\sim200$ km further. At this point, $\rm T_{R}$ gradually rises, similar to the chromospheric behavior in models obtained by \citet{Gouttebroze} and \citet{ferrero95}. In Fig.\, \ref{kp-altair}(c) we plot the contribution function \citep[CF;][]{tapia20} to constrain where the emission is generated.

Previous modeling approaches primarily considered Altair's UV spectrum to compute the structure of the upper photosphere and chromosphere. Anomalies in the CII line were speculated to be from ISM absorption \citep{Gouttebroze}. Our model could not produce the observed 1.3 mm $\rm T_{B}$ minimum and therefore we did not include these data in the final KP or PHOENIX results. If there is significant foreground absorption, it isn't clear if this feature is observable at 1.3 mm. Absorption from interstellar grains would yield a broad feature with a spectral index reflective of the grain's size distribution \citep[e.g.,][]{jones96}. At 1.3 mm, Altair's emission would be predominately absorbed by $\sim$mm-sized grains, and ISM grains are typically a few $\mu$m in size \citep[e.g.,][]{weingartner01}. If the absorption is from a molecular line, e.g. CO(2--1) with a rest frequency of 230.53 GHz (1.3 mm), then the approximately flat spectral index between the NOEMA sidebands of 1.29 mm and 1.38 mm implies a line width $\geq0.1$ mm and a velocity of $\sim10^{4}~\rm km~s^{-1}$, which is unphysical for the ISM. We determine that the 1.3 mm data are likely not contaminated by foreground material. 

The observed 1.3 mm flux density is broadly consistent between the two NOEMA semesters. After consultation with NOEMA support scientists we conclude that significant instrument or calibration uncertainties are unlikely. While there is no known physical mechanism that could reproduce the 1.3 mm flux, we cannot rule out an anomalous absorption or opacity in Altair's atmosphere without more data. Future observations at $0.25-1.3$ mm, and follow-up observations at 1.3 mm with another facility such as ALMA, are therefore imperative to accurately determine the cause of the observed drop in flux.

\subsection{Applicability to observations of debris disks}\label{sec:disks}

The radio spectra obtained with The MESAS Project are useful outside the context of stellar atmospheric processes. Historically, the presence of debris disks are inferred by excess over the expected stellar emission at a given wavelength when the disks cannot be spatially resolved \citep{aumann84}. The first star observed in the campaign was Sirius A, a star with no known circumstellar debris \citep{white18}. Radio observations of Sirius A were used to inform a PHOENIX model atmosphere. This spectrum was applied to the millimeter emission of Fomalhaut, an A3V star very similar to Sirius A but with an extended debris disk \citep[e.g.,][]{kalas05,acke12,white17}. In addition to Fomalhaut's well-resolved debris ring, there is an inferred inner asteroid belt that has not been resolved \citep{acke12}. Such an asteroid belt would be collisionally evolving to produce millimeter-sized debris that may be detectable at millimeter wavelengths. Instead, Fomalhaut's millimeter emission has a spectrum nearly identical to that of Sirius A \citep{su16,white17}, which has no known debris. This result casts doubt on the presence of an additional debris component in the Fomalhaut system \citep{white_thesis} and highlights the importance of an accurate radio stellar spectrum to determine the presence and characteristic of unresolved circumstellar debris. 

\subsection{Future Work}

As is clear from Altair's observed radio spectrum, determining the stellar flux at a given wavelength based solely on the flux at a different wavelength is non-trivial. In order to fully characterize the radio spectrum of Altair, observations between $0.25-1.29$ mm are imperative. The only facility that can fill in the gap on Altair's missing data is ALMA. Follow-up observations at all wavelengths with multiple cadences will assess the long term variability or stability of Altair's radio emission.

The MESAS Project is currently pushing state-of-the-art radio observatories to their limits. Proposed future facilities, such as the next generation Very Large Array (ngVLA), will be invaluable for revealing stellar atmosphere structures. For example, the ngVLA will yield a 93 GHz (3.22 mm) continuum $ \rm \sigma_{\rm RMS}  = 0.40 - 1.0 ~\mu Jy~beam^{-1}$ in 1 hour\footnote{For estimated ngVLA sensitivities, see http://library.nrao.edu/public/memos/ngvla/NGVLA\_21.pdf}. In addition, a majority of the proposed ngVLA antennas will have a maximum baseline of 1000 km, giving a resolution of $0.66$ mas. The ngVLA will resolve Altair by $5-6$ beam widths in only a few minutes of integration time. This level of detail is absolutely imperative to determine how the high rotational velocity, and resulting dynamo, can impact the overall $\rm T_{B}$ structure of the star. The ngVLA’s unprecedented sensitivity means all G-type and brighter stars are detectable within $\sim75$ pc, allowing for a comprehensive survey of our Solar neighborhood \citep[see, e.g.,][]{white_ngvla, carilli18}. 

\section{Conclusions} \label{sec:conclusion}

We present the first observationally informed sub-millimeter to centimeter spectrum of a rapidly rotating A-type star. We observed Altair with NOEMA and VLA at $1.29 - 9.1$ mm in 2018 and 2019. The data were used to inform both PHOENIX and KINICH-PAKAL stellar atmosphere modeling codes. We conclude that the KINICH-PAKAL model better reproduces the data. 
This temperature spectrum is cooler than the photosphere at $<1$ mm then becomes much hotter at $>3$ mm, indicative of a chromosphere. The radio spectrum deviates from that expected for both A-type stars and cooler F/G-type stars. In addition, the significantly smaller flux observed at 1.3 mm may be due to an anomalous opacity which we were unable to reproduce in our models. These results are part of The MESAS Project, an ongoing observational campaign that seeks to build a radio catalog of main-sequence stars.

We thank the referee for feedback that improved the manuscript. The National Radio Astronomy Observatory is a facility of the National Science Foundation operated under cooperative agreement by Associated Universities, Inc. This work is based on observations carried out under project numbers W17BF and W18BQ with the IRAM NOEMA Interferometer. IRAM is supported by INSU/CNRS (France), MPG (Germany) and IGN (Spain). This research has made use of the CIRADA cutout service at URL cutouts.cirada.ca, operated by the Canadian Initiative for Radio Astronomy Data Analysis (CIRADA). CIRADA is funded by a grant from the Canada Foundation for Innovation 2017 Innovation Fund (Project 35999), as well as by the Provinces of Ontario, British Columbia, Alberta, Manitoba and Quebec, in collaboration with the National Research Council of Canada, the US National Radio Astronomy Observatory and Australia’s Commonwealth Scientific and Industrial Research Organisation.

\vspace{5mm}
\facilities{NOEMA, VLA}

\software{{\scriptsize CASA 5.4.1} \citep{casa_reference}; PHOENIX version 17.01.02A \citep{hauschildt99} 
          }





\bibliographystyle{aasjournal}

\begin{thebibliography}{}

\bibitem[Abt \& Morrell(1995)]{abt95}Abt, H.A., \& Morrell, N.I., 1995, ApJS, 99, 135

\bibitem[Acke et al.(2012)]{acke12}Acke, B., Min, M., Dominik, C., et al., 2012, A\&A, 540, A125

\bibitem[Aufdenberg et al.(2006)]{aufdenberg06} Aufdenberg, J.P., M\'erand, A., Du Foresto, V.C., et al., 2006, ApJ, 645(1), 664

\bibitem[Aumann et al.(1984)]{aumann84}Aumann, H. H., Gillett, F. C., Beichman, C. A., et al. 1984, ApJ, 278, L23

\bibitem[Avrett \& Loeser(2008)]{avrett08}Avrett, E. H., \& Loeser, R. 2008, ApJS, 175, 229

\bibitem[Ayres(2010)]{ayres} Ayres, T.\ 2010, \memsai, 81, 553

\bibitem[Bouchaud et al.(2020)]{bouchaud20}Bouchaud, K., de Souza, A.D., Rieutord, M., et al., 2020, A\&A, 633, A78

\bibitem[Buzasi et al.(2005)]{buzasi05}Buzasi, D.L., Bruntt, H., Bedding, T.R., et. al., 2005, ApJ, 619(2), 1072

\bibitem[Carilli et al.(2018)]{carilli18}Carilli, C.L., Butler, B., Golap, K., et al., 2018, in ASP Conf. Ser. 517,
Science with a Next Generation Very Large Array, ed. E. Murphy, 369

\bibitem[Chabrier \& K\"uker(2006)]{chabrier06}Chabrier, G. and K\"uker, M., 2006, A\&A, 446(3), 1027

\bibitem[Charbonneau(2013)]{charbonneau13}Charbonneau, P., 2013, in Solar and Stellar Dynamos, Springer, Berlin, Heidelberg, 187

\bibitem[De la Luz et al.(2011)]{delaluz2011} De la Luz, V., Lara, A., \& Raulin, J.-P.\ 2011, \apj, 737, 1

\bibitem[Ferrero et al.(1995)]{ferrero95}Ferrero, R.F., Gouttebroze, P., Catalano, S., et al., 1995, ApJ, 439, 1011

\bibitem[Gouttebroze et al.(1999)]{Gouttebroze} Gouttebroze, P., Ferrero, R.~F., Marilli, E., et al.\ 1999, \aap, 348, 198

\bibitem[Guerrero et al.(2016)]{guerrero16}Guerrero, G., Smolarkiewicz, P.K., Dal Pino, E.D.G., et al., 2016, ApJ, 819(2), 104.

\bibitem[Hauschildt \& Baron(1999)]{hauschildt99}Hauschildt, P. H. \& Baron, E. 1999, Journal of Comp. and App. Math, 109, 41

\bibitem[Hughes et al.(2019)]{hughes19}Hughes, A.G., Boley, A.C., Osten, R.A., et al., 2019, ApJ, 881(1), 33

\bibitem[Jones et al.(1996)]{jones96}Jones, A.P., Tielens, A.G.G.M., \& Hollenbach, D.J., 1996, ApJ, 469, 740

\bibitem[Kalas et al.(2005)]{kalas05}Kalas, P., Graham, J. R., \& Clampin, M., 2005, Nature, 435(7045), 1067

\bibitem[Linsky(2017)]{linsky17} Linsky, J.~L.\ 2017, \araa, 55, 15

\bibitem[Lipatov \& Brandt(2020)]{lipatov20}Lipatov, M. \& Brandt, T.D., 2020, ApJ, 901, 100

\bibitem[Liseau et al.(2013)]{liseau13} Liseau, R., Montesinos, B., Olofsson, G., et al.\ 2013, \aap, 549, L7

\bibitem[Liseau et al.(2016)]{Liseau} Liseau, R., De la Luz, V., O'Gorman, E., et al.\ 2016, \aap, 594, A109

\bibitem[Loukitcheva et al.(2004)]{loukitcheva04}Loukitcheva, M., Solanki, S. K., Carlsson, M., et al. 2004, A\&A, 419, 747

\bibitem[McMullin et al.(2007)]{casa_reference}McMullin J.P., Waters B., Schiebel D., et al., 2007, Astronomical Data Analysis Software and Systems XVI (ASP Conf. Ser. 376), ed. R. A. Shaw, F. Hill, \& D. J. Bell (San Francisco, CA: ASP), 127

\bibitem[Monnier et al.(2007)]{monnier07}Monnier, J.D., Zhao, M., Pedretti, E., et al., 2007, Science, 317(5836), 342

\bibitem[Preston(1974)]{preston74}Preston, G.W. 1974, ARA\&A, 12, 257

\bibitem[Robrade \& Schmitt(2009)]{robrade09}Robrade, J. \& Schmitt, J.H.M.M., 2009, A\&A, 497(2), 511-520

\bibitem[Rodr\'iguez et al.(2003)]{rodriguez03}Rodr\'iguez, L.F., Porras, A., Claussen, M.J., et al., 2003, ApJL, 586(2), L137

\bibitem[Schmitt et al.(1985)]{schmitt85}Schmitt, J.H.M.M., Golub, L., Harnden Jr, F.R., et al., 1985, ApJ, 290, 307

\bibitem[Schulz et al.(2017)]{schulz17}Schulz, B., Marton, G., Valtchanov, I., et al., 2017, SPIRE Point Source Catalog Explanatory Supplement, arXiv:1706.00448

\bibitem[Shorlin et al.(2002)]{shorlin02}Shorlin, S.L.S., Wade, G.A., Donati, J.F., et al., 2002, A\&A, 392(2), 637

\bibitem[Simon et al.(1995)]{simon95}Simon, T., Drake, S.A., \& Kim, P.D., 1995, ASP, 107(717), 1034

\bibitem[Su et al.(2016)]{su16}Su, K. Y., Rieke, G. H., Defr\'ere, D., et al. 2016, ApJ, 818, 45

\bibitem[Takeda(2020)]{takeda20}Takeda, Y., 2020, MNRAS, 499(1), 1126

\bibitem[Tapia-V\'azquez \& De la Luz(2020)]{tapia20}Tapia-V\'azquez, F., \& De la Luz, V. 2020, ApJS, 246, 5

\bibitem[Thureau et al.(2014)]{thureau14}Thureau, N.D., Greaves, J.S., Matthews, B.C., et al., 2014, MNRAS, 445(3), 2558

\bibitem[Trigilio et al.(2018)]{trigilio} Trigilio, C., Umana, G., Cavallaro, F., et al.\ 2018, \mnras, 481, 217

\bibitem[Villadsen et al.(2014)]{villadsen} Villadsen, J., Hallinan, G., Bourke, S., et al.\ 2014, \apj, 788, 112

\bibitem[van Leeuwen(2007)]{leeuwen07}van Leeuwen, F., 2007, A\&A, 474(2), pp.653-664

\bibitem[Virtanen et al.(2020)]{virtanen20}Virtanen, P., Gommers, R., Oliphant, T.E., et al. 2020, Nature methods, 17(3), 261

\bibitem[Walter et al.(1995)]{walter95}Walter, F.M., Matthews, L.D. and Linsky, J.L., 1995, ApJ, 447, 353

\bibitem[Weingartner \& Draine(2001)]{weingartner01}Weingartner, J.C. \& Draine, B.T., 2001, ApJ, 548(1), 296

\bibitem[White et al.(2017)]{white17}White, J. A., Boley, A. C., Dent, W. R. F., et al. 2017a, MNRAS, 466, 4201

\bibitem[White et al.(2018a)]{white18} White, J.A., Aufdenberg, J., Boley, A.C., et al., 2018a, ApJ, 859(2), 102

\bibitem[White(2018b)]{white_thesis}White, J. A. 2018, Doctoral dissertation, Univ. British Columbia

\bibitem[White et al.(2018c)]{white_ngvla} White, J.A., Aufdenberg, J., Boley, A.C., et al., 2018b, in ASP Conf. Ser. 517,
Science with a Next Generation Very Large Array, ed. E. Murphy, 171

\bibitem[White et al.(2019)]{white19}White, J. A., Aufdenberg, J., Boley, A. C., et al. 2019, ApJ, 875, 55

\bibitem[White et al.(2020)]{white20} White, J.A., Tapia-V\'azquez, F., Hughes, A.G., et al., 2020, ApJ, 894(1), 76

\end{thebibliography}

\appendix

\section{Observational Data}\label{data_app}

\subsection{NOEMA}\label{sec:noema}

Altair was observed with NOEMA in 2018 on April 28, May 12, and May 17 (ID W17BF, PI White), and in 2019 on February, 07, 08, and 15 (ID W18BQ, PI White). The observations were centered on Altair using J2000 coordinates RA = 19$^{\rm hr}$ 50$^{\rm  min}$ 46.996$^{\rm  s}$ and $\delta = +08^{\circ} ~52' ~05\farcs956$ and accounting for proper motion. The observations in 2018 used 9 antennas in the C and D configurations with baselines ranging from $24-293$ m and $24-176$ m, respectively. The observations in 2019 used 10 antennas in the A configuration with baselines ranging from $32-760$ m.

The observations used the Band 1, Band 2, and Band 3 instruments with the PolyFiX correlator and 4064 x 2 MHz channels in each band. In Band 1, the rest frequency of each baseband was tuned to 214.208 GHz, 218.269 GHz, 229.693 GHz, and 233.754 GHz. In Band 2, they were tuned to 134.215 GHz, 138.276 GHz, 149.700 GHz, and 153.761 GHz. In Band 3, they were tuned to 84.219 GHz, 88.280 GHz, 99.704 GHz, 103.766 GHz. An identical instrument setup was used in 2018 and 2019. The data calibration pipeline and imaging utilized the {\scriptsize  GILDAS} software packages and was done with the assistance of NOEMA contact scientists at IRAM. In 2018, quasars J2002+150 and 1932+204 were used for phase and amplitude calibration and MWC349 was used to calibrate the absolute flux. The Band 1 and Band 2 observations used 3C345 for bandpass calibration and 3C454.3 was used for Band 3. In 2019, all bands used quasars J2002+150 and J1938+048 for phase and amplitude calibrations and MWC349 for absolute flux calibration. Quasars 1633+382, 1749+096, and 3C345 were used as bandpass calibrators in Band 1, Band 2, and Band 3, respectively. 

In each of the 3 receivers, the 2 upper sidebands were combined and the 2 lower sidebands were combined to increase the signal-to-noise. This gives effective wavelengths of 1.29 mm, 1.38 mm, 1.98 mm, 2.20 mm, 2.96 mm, and 3.48 mm for each data point. To derive the flux at each wavelength, we used the GILDAS MAPPING package and the \textit{UVFIT} task to fit a point source model directly to the visibilities (see Table\,\ref{data_summary} for a list of fluxes). We used the \textit{UVMAP} task to make dirty image maps and then Hogbom cleaning to a threshold of 1/2 the RMS noise. The $\sigma_{rms}$ of the images is adopted as the representative uncertainty for each data point. The absolute flux calibration uncertainty for NOEMA is $\sim20\%$ at 1 mm and less than $10\%$ at 3 mm\footnote{NOEMA absolute flux calibration uncertainties https://www.iram.fr/IRAMFR/GILDAS/doc/html/pdbi-cookbook-html/pdbi-cookbook.html}. 

In 2018, from shortest to longest wavelength, the observations achieve sensitivities and synthesized beams of: $107~\rm \mu Jy~beam^{-1}$ and $2\farcs68\times 1\farcs60$ at a position angle (PA) of $10.5^{\circ}$; $83.9~\rm \mu Jy~beam^{-1}$ and $2\farcs85\times 1\farcs68$ at a PA of $5.9^{\circ}$; $45.0~\rm \mu Jy~beam^{-1}$ and $3\farcs44\times 2\farcs11$ at a PA of $6.9^{\circ}$; $32.8~\rm \mu Jy~beam^{-1}$ and $3\farcs82\times 2\farcs27$ at a PA of $7.5^{\circ}$; $25.1~\rm \mu Jy~beam^{-1}$ and $2\farcs78\times 1\farcs64$ at a PA of $31.5^{\circ}$; and $29.8~\rm \mu Jy~beam^{-1}$ and $3\farcs18\times 1\farcs85$ at a PA of $40.0^{\circ}$, respectively. 

In 2019, the observations achieve sensitivities and synthesized beams of: $75.1~\rm \mu Jy~beam^{-1}$ and $0\farcs76\times 0\farcs37$ at a PA of $28.4^{\circ}$; $66.0~\rm \mu Jy~beam^{-1}$ and $0\farcs79\times 0\farcs39$ at a PA of $28.9^{\circ}$; $45.6~\rm \mu Jy~beam^{-1}$ and $1\farcs14\times 0\farcs53$ at a PA of $17.2^{\circ}$; $36.7~\rm \mu Jy~beam^{-1}$ and $1\farcs18\times 0\farcs58$ at a PA of $18.7^{\circ}$; $30.8~\rm \mu Jy~beam^{-1}$ and $1\farcs64\times 0\farcs75$ at a PA of $17.2^{\circ}$; and $27.7~\rm \mu Jy~beam^{-1}$ and $1\farcs84\times 0\farcs86$ at a PA of $18.0^{\circ}$, respectively.

\subsection{VLA} \label{sec:vla}

The data from the VLA were acquired during 3 Scheduling Blocks (SBs) on 2019 March 15, March 19, and March 24 (ID 19A-141, PI White). The observations were centered on Altair using J2000 coordinates RA = 19$^{\rm hr}$ 50$^{\rm  min}$ 46.996$^{\rm  s}$ and $\delta = +08^{\circ} ~52' ~05\farcs956$ and accounted for proper motion. The data used 27 antennas in the B configuration with baselines ranging from 0.21 to 11.1 km.

Each SB used an identical instrument configuration and observing strategy. The 33 GHz data utilized the Ka Band tuning setup with $4\times2.048$ GHz basebands and rest frequency centers of 28.976 GHz, 31.024 GHz, 34.976 GHz, and 37.024 GHz, yielding an effective frequency of 33 GHz (9.1 mm). The 45 GHz data utilized the Q Band tuning setup with $4\times2.048$ GHz basebands and rest frequency centers of 41.024 GHz, 43.072 GHz, 46.968 GHz, and 48.976 GHz, yielding an effective frequency of 45 GHz (6.7 mm).

The SBs were reduced using the {\scriptsize CASA 5.4.0} pipeline \citep{casa_reference}, which included bandpass, flux, and phase calibrations. For all SBs, Quasar J1950+0807 was used for phase and bandpass calibrations. Quasar 0137+331=3C48 was used as a flux calibration source for all SBs. 3C48 was originally intended to also be used as a bandpass calibrator, but a low signal-to-noise in each SPW made it unreliable. Upon consultation with the VLA HelpDesk, we modified the scan intents of the raw data to identify J1950+0807 as the bandpass calibrator and proceeded with a standard pipeline calibration. The on-going flare of the calibrator increased absolute flux uncertainty to $\sim20\%$ in the Q band and $\sim15\%$ in the Ka Band\footnote{For a note on the VLA flux calibration uncertainty, see science.nrao.edu/facilities/vla/docs/manuals/oss/performance/fdscale. }.

The three SBs were concatenated using the {\scriptsize CASA} task \textit{concat}. To derive the flux at each wavelength, we used the {\scriptsize CASA} task \textit{uvmodelfit} to fit a point source model at each frequency. The concatenated data were imaged with a natural weighting and cleaned using {\scriptsize CASA}'s \textit{CLEAN} algorithm down to a threshold of 1/2 the RMS noise. The observations achieve a sensitivity of $21.9~\rm \mu Jy~beam^{-1}$ and $5.9~\rm \mu Jy~beam^{-1}$ for the 6.7 mm and 9.0 mm data, respectively. The size of the resulting synthesized beam is $0\farcs22\times 0\farcs17$ at a position angle of $-32.4^{\circ}$ and $0\farcs25\times 0\farcs21$ at a position angle of $-21.6^{\circ}$, respectively.

\subsection{Literature data} \label{sec:lit}

In addition to the new data presented here, we also include \textit{Herschel} and VLA Sky Survey (VLASS) data from literature in our analysis. Altair was observed with the \textit{Herschel} PACS instrument at 0.10 mm and 0.16 mm on 2010 May 09. We use the two IR photometry data points of Altair presented in \citet{thureau14} as part of the DEBRIS survey (we note that the authors report no excess flux that could be attributed to circumstellar material). The observed flux was $329.02\pm18.88$ mJy and $148.26\pm9.94$ mJy at 0.10 mm and 0.16 mm, respectively. We also use \textit{Herschel} SPIRE data from the \textit{Herschel} Point Source Catalog \citep{schulz17}. Altair was observed with SPIRE on 2011 NOV 16. The observed flux was $55.7\pm8.3$ mJy at 0.25 mm.

VLASS is an all sky survey with uniform sensitivity at 3 GHz (100 mm) covering 3 separate epochs. At the time of writing, only one epoch had been completed. We downloaded a $1'$ section of the sky centered on the J2000 coordinates of Altair from The Canadian Initiative for Radio Astronomy Data Analysis (CIRADA) server\footnote{CIRADA cutout service at URL cutouts.cirada.ca}. The observations were made on 2017 October 25 and have a synthesized beam size of $2\farcs68 \times 2\farcs32$ and a position angle of $25.5^{\circ}$. Altair was not detected, but we use the $\sigma_{\rm RMS}= 120~\rm \mu Jy~beam^{-1}$ from the surrounding area as an upper level limit on the 3 GHz flux.

\begin{table*}

\centering 
\begin{tabular}{c  c  c  c  c   c  c c } 
\hline\hline \ 
   	Wavelength  & Facility & Date & Flux Calibrator  & Beam Size & Beam PA & Flux Density & Uncertainty \\
   		$($mm) &  & YYYY MMM DD & & ($''$) & ($^{\circ}$) & ($\rm \mu Jy$) &  ($\rm \mu Jy$)  \\
   	\hline 
   	0.10 & Herschel$^{a}$ & 2010 MAY 09  & - & - & - & 329020  & 18880  \\
   	0.16 & Herschel$^{a}$ & 2010 MAY 09  & - & - & - & 148260  & 9940  \\
   	0.25 & Herschel$^{b}$ & 2011 NOV 16  & - & - & - & 55700  & 8300  \\
	1.29 & NOEMA   & 2018 MAY 17  & MWC349$^{c}$ & $2\farcs68\times1\farcs60$ & 10.5 & 431  & 107  \\
	1.38 & NOEMA   & 2018 MAY 17  & MWC349 & $2\farcs85\times1\farcs68$ & 5.9  & 548  & 83.9  \\
	1.98 & NOEMA   & 2018 MAY 12  & MWC349 & $3\farcs44\times2\farcs11$ & 6.9  & 705  & 45.0  \\
	2.20 & NOEMA   & 2018 MAY 12  & MWC349 & $3\farcs82\times2\farcs27$ & 7.5  & 668  & 32.8  \\
	2.96 & NOEMA   & 2018 APR 28  & MWC349 & $2\farcs78\times1\farcs64$ & 31.5 & 384  & 25.1  \\
	3.48 & NOEMA   & 2018 APR 28  & MWC349 & $3\farcs18\times1\farcs85$ & 40.0 & 283  & 29.8  \\
    1.29 & NOEMA   & 2019 FEB 07  & MWC349 & $0\farcs76\times0\farcs37$ & 28.4 & 395  & 75.1  \\
	1.38 & NOEMA   & 2019 FEB 07  & MWC349 & $0\farcs79\times0\farcs39$ & 28.9 & 342  & 66.0  \\
	1.98 & NOEMA   & 2019 FEB 08  & MWC349 & $1\farcs14\times0\farcs53$ & 17.2 & 596  & 45.6  \\
	2.20 & NOEMA   & 2019 FEB 08  & MWC349 & $1\farcs18\times0\farcs58$ & 18.7 & 469  & 36.7  \\
	2.96 & NOEMA   & 2019 FEB 15  & MWC349 & $1\farcs64\times0\farcs75$  & 17.2 & 286  & 30.8  \\
	3.48 & NOEMA   & 2019 FEB 15  & MWC349 & $1\farcs84\times0\farcs86$  & 18.0 & 262  & 27.7  \\
	6.7  & VLA     & 2019 MAR 15, 19, 24 & 3C48$^{d}$ & $0\farcs22\times0\farcs17$ & $-32.4$ & 83 & 21.9 \\
	9.1  & VLA     & 2019 MAR 15, 19, 24 & 3C48$^{d}$ & $0\farcs25\times0\farcs21$ & $-21.6$ & 57 & 5.9 \\
    100  & VLASS$^{e}$   & 2017 OCT 25 & - &  $2\farcs68 \times 2\farcs32$  & 25.5 & $<120^{f}$ & - \\
	
\hline	
    
\end{tabular}
\caption{Summary of the new observations presented here and data from the literature. The NOEMA and VLA data use a best fit flux from modeling the visibility data as described in Sections\,\ref{sec:noema} and \ref{sec:vla}, respectively. The $\sigma_{\rm RMS}$ uncertainties are from the CLEANed images and do not include an absolute flux calibration uncertainty. The calibration uncertainties are $\sim20\%$ at 1 mm and $10\%$ at 3 mm for NOEMA, and $\sim20\%$ at 6.7 mm and $\sim15\%$ at 9.1 mm for the VLA. (a) \textit{Herschel} PACS photometry data \citep{thureau14}. (b) Photometry data from the \textit{Herschel} SPIRE Point Source Catalog \citep{schulz17}. (c) The flux calibrator MWC349 for NOEMA was observed in each of the tracks and its model flux was used to derive the flux of the calibrators and the source. (d) The quasar 3C48, which was used to calibrate the flux from the VLA observations, has been undergoing a flare since early 2018 and therefore contributes to a larger absolute flux calibration uncertainty than is typical for these wavelength data (see Sec.\ref{sec:vla}). (e) VLASS data obtained the CIRADA server. (f) Altair was not detected in the VLASS data, therefore the $\sigma_{\rm RMS}$ of the area around Altair's expected location is taken as an upper level limit to the flux.}

\label{data_summary}
\end{table*}



\end{document}